\documentstyle[prd,aps,floats]{revtex}
\begin{document}
\draft
%
%
\input epsf
\renewcommand{\topfraction}{0.8}
\twocolumn[\hsize\textwidth\columnwidth\hsize\csname
@twocolumnfalse\endcsname
\preprint{CERN-TH/97-329, SU-ITP-97-49, hep-ph/9711360}
\title{Preheating in Hybrid Inflation}
\author{Juan Garc\'{\i}a-Bellido}
\address{Theory Division, CERN, CH-1211 Geneva 23, Switzerland}
\author{Andrei Linde}
\address{Physics Department, Stanford University,
Stanford CA 94305-4060, USA}
\date{November 17, 1997}
\maketitle
\begin{abstract}

  We investigate the possibility of preheating in hybrid inflation.
  This scenario involves at least two scalar fields, the inflaton
  field $\phi$, and the symmetry breaking field $\sigma$. We found
  that the behavior of these fields after inflation, as well as the
  possibility of preheating (particle production due to parametric
  resonance), depends crucially on the ratio of the coupling constant
  $\lambda$ (self-interaction of the field $\sigma$) to the coupling
  constant $g^2$ (interaction of $\phi$ and $\sigma$). For $\lambda
  \gg g^2$, the oscillations of the field $\sigma$ soon after
  inflation become very small, and all the energy is concentrated in
  the oscillating field $\phi$.  For $\lambda \sim g^2$ both fields
  $\sigma$ and $\phi$ oscillate in a rather chaotic way, but
  eventually their motion stabilizes, and parametric resonance with
  production of $\chi$ particles becomes possible. For $\lambda \ll
  g^2$ the oscillations of the field $\phi$ soon after inflation
  become very small, and all the energy is concentrated in the
  oscillating field $\sigma$. Preheating can be very efficient if the
  effective masses of the fields $\phi$ and $\sigma$ are much greater
  than the Hubble constant at the end of inflation, since those fields
  can then oscillate many times per $e$-fold, with a large amplitude. 
  Preheating can also be efficient if these fields are coupled to
  other light scalar (or vector) fields~$\chi$. In the recently
  proposed hybrid models with a second stage of inflation after the
  phase transition, both preheating and usual reheating are
  inefficient.  Therefore for a very long time the universe remains in
  a state with vanishing pressure. As a result, density contrasts
  generated during the phase transition in these models can grow and
  collapse to form primordial black holes. Under certain conditions,
  most of the energy density after inflation will be stored in small
  black holes, which will later evaporate and reheat the universe.

\end{abstract}
\pacs{PACS:~ 98.80.Cq \hspace{1.6cm}  CERN-TH/97-329,~~~
SU-ITP-97-49 \hspace{1.6cm} hep-ph/9711360}

\vskip2pc]

\section{Introduction}

In most inflationary models the energy density at the end of inflation
is released when the scalar field oscillates near the minimum of its
potential, creates elementary particles, which later collide, decay
into other particles, and eventually thermalize via various
interactions ~\cite{book}. The original version of the theory of
reheating of the universe was based on the idea that the oscillating
scalar field can be considered as a collection of scalar particles,
each of which decays independently ~\cite{reheat}. This simple
picture, based on perturbation theory, is indeed valid in many
inflationary models.  However, recently it was shown that in many
versions of chaotic inflation there is an additional stage when the
oscillating scalar field produces other fields in a regime of
parametric resonance.  This regime was called
``preheating''~\cite{KLS1}. The process is nonperturbative and in
certain cases it can be extremely efficient. It may change the final
value of the temperature of the universe after
reheating~\cite{KLS1,BB}, it may lead to specific nonthermal phase
transitions~\cite{PhaseTr}, and it may provide a new mechanism for the
generation of the baryon asymmetry~\cite{KolbRiotto}.  For a detailed
discussion of the theory of preheating see~\cite{KLS2,GKLS} and
references therein.

In this paper we will discuss preheating in the context of hybrid
inflation models~\cite{hybrid}.  These models became rather popular
lately because they can be relatively easily formulated in the context
of supersymmetric theories. They allow inflation to occur at very
small values of scalar fields, and at a relatively small energy
density, which removes some restrictions on inflationary models that
previously seemed almost unavoidable.  This scenario is also interesting
from the point of view of microwave background anisotropies and large
scale structure because it provides natural models for tilted
primordial spectra of density perturbations~\cite{GBL}.

Hybrid inflation models describe at least two classical scalar fields,
$\phi$ and $\sigma$. After the end of inflation both of these fields
oscillate, and the energy of their oscillations can be transferred to
the production of $\phi$ and $\sigma$ particles, or to the energy of
other particles interacting with these fields.

The investigation of reheating in these models should consist of
several steps.  First of all, one should study the oscillations of the
fields $\phi$ and $\sigma$ after inflation, neglecting particle
production.  Until now, this important question was only partially
analysed in some special situations. As we will show, the behavior of
these two fields after inflation crucially depends on the relation
between coupling constants in the theory. In some cases soon after
inflation one of these fields rapidly relaxes near the minimum of the
effective potential, whereas the other one continues to oscillate.
Then all the energy after inflation is concentrated in this one field,
and reheating occurs due to its decay. We will show how to find which
of the two field will play this role. In some other cases, the motion
of these two fields becomes chaotic. Both of these fields oscillate
simultaneously, and their energy transfers from one field to another
many times during these oscillations. Without investigation of this
issue one cannot achieve a proper understanding of reheating in hybrid
inflation.

The second issue to study is the possibility of a parametric resonance
which would transfer energy of the homogeneous oscillating classical
fields $\phi$ and $\sigma$ to the energy of elementary particles
$\phi$ and $\sigma$.  As we will see, this process in general is
possible, but typically it is not very efficient.  Then one should
analyse a possibility of a resonant decay of the fields $\phi$ and
$\sigma$ to some other scalar or vector particles.

When this investigation is completed, one should study other channels
of decay of the classical fields $\phi$ and $\sigma$, such as the
usual (non-resonant) decay to their own quanta and into other
particles in accordance to the elementary theory of
reheating~\cite{reheat}.  Indeed, as it was shown in Ref.~\cite{KLS2},
this is a separate process, which differs from the resonant
decay in the limit when parametric resonance becomes inefficient.

Until now, investigation of reheating in hybrid inflation concentrated
on the non-resonant decay of the fields $\phi$ and $\sigma$, see
e.g.~\cite{Lazar}.  The results obtained by this method may be quite
correct in some models where there is no parametric resonance, but if
parametric resonance is possible, then the theory of reheating becomes
completely different.

The purpose of this paper is to investigate this issue.  Our paper is
not going to be a complete study of preheating in hybrid inflation. We
will restrict ourselves to the investigation of the first stages of
preheating in hybrid inflation, which is sufficient to identify the
models where preheating may occur.  A complete investigation of
preheating in hybrid inflation can be achieved using the analytical
methods developed in~\cite{KLS2}, together with numerical lattice
simulations along the lines of Refs.~\cite{Tkachev}.

The main part of our paper will be devoted to the standard version of
hybrid inflation with one stage of inflation. One may also
consider inflationary models with two stages of inflation~\cite{Guth}.
We will show that in such models both preheating and usual reheating
are inefficient. However, under some conditions reheating in such
models may be rather efficient, but it occurs in a rather unusual way.
In a recent paper~\cite{GBLW} we have shown that large density
perturbations could be produced in such models at the time
corresponding to the beginning of the second stage of inflation. In
that case, large density contrasts re-enter the horizon during the
radiation era and can collapse to form black holes.  Radiation
pressure may prevent the formation of primordial black holes, unless
their density contrast is of order one. However, if the process of
reheating after inflation is very inefficient, then pressure after
inflation will be negligible, and nothing will prevent the formation
of primordial black holes. If these black holes have a mass of order
$10^{-5} - 10^{-9}$~g, they will evaporate very early and reheat the
universe~\cite{GBLW}, producing all the entropy we observe today. This
is a very interesting possibility which needs to be further explored.

\section{The stage of hybrid inflation}\label{DYN}

The simplest realization of chaotic hybrid inflation is provided by
the potential~\cite{hybrid}
\begin{equation}\label{hybrid}
V(\phi,\sigma) = {1\over4\lambda}\left(M^2-\lambda\sigma^2 \right)^2
 + {1\over2}m^2\phi^2 + {1\over2}g^2\phi^2\sigma^2 \, .
\end{equation}
The bare masses $m$ and $M$ of the scalar fields $\phi$ and $\sigma$
are ``dressed'' by
their mutual interaction. At large values of the fields, their effective
masses squared are both positive and the potential has the symmetry
$\sigma \leftrightarrow -\sigma$. At small values of the field $\phi$,
the potential has a maximum at $\phi=\sigma=0$ and a
global minimum at $\phi=0, \sigma=\sigma_0\equiv M/\sqrt\lambda$,
where the above symmetry is broken.

The complete equations of motion for the homogeneous fields are
\begin{eqnarray}\label{EQM}
\ddot\phi + 3H\dot\phi
 &=& - (m^2 +g^2\sigma^2) \phi\, , \\
\ddot\sigma + 3H\dot\sigma
 &=& (M^2 - g^2\phi^2 - \lambda\sigma^2) \sigma \, ,
\end{eqnarray}
subject to the Friedmann constraint,
\begin{equation}
H^2 = {8\pi\over3M^2_{\rm P}} \Big[ {1\over2}\dot\phi^2 +
{1\over2}\dot\sigma^2 + V(\phi,\sigma)\Big] \, .
\end{equation}
Motion starts at large $\phi$, where the effective mass of the $\sigma$
field is large and the field is sitting at the minimum of the potential
at $\sigma=0$. As the field $\phi$ decreases, its quantum fluctuations
produce an almost scale invariant but slightly tilted spectrum
of density perturbations~\cite{hybrid,LL93,GBW}.

During the slow-roll of the field $\phi$, the effective mass of the
triggering field is $m^2_\sigma = g^2\phi^2 - M^2$. When the field
$\phi$ acquires the critical value $\phi_c \equiv M/g$, fluctuations
of the massless $\sigma$ field trigger the symmetry breaking phase
transition that ends inflation. If the bare mass $M$ of the $\sigma$
field is large compared with the rate of expansion $H$ of the
universe, the transition will be instantaneous and inflation will end
abruptly, as in the original hybrid inflation
model~\cite{hybrid,CLLSW}. If on the contrary the bare mass $M$ is of
the order of $H$, then the transition will be very slow and there is
a possibility of having yet a few more $e$-folds of inflation after
the phase transition, see Refs.~\cite{Guth,GBLW}.

When $\sigma=0$ the inflaton potential becomes $V(\phi) = M^4/4\lambda
+ m^2\phi^2/2$. Since the scalar field $\phi$ is of the order of
$\phi_c = M/g$, for $m^2\ll g^2M^2/\lambda$ the energy density is
dominated by the vacuum energy,
\begin{equation}\label{H0}
H^2_0 = {2\pi\over3\lambda}\,{M^4\over M^2_{\rm P}} \, ,
\end{equation}
to very good accuracy~\cite{GBW}. At that time $\sigma = 0$, and one can
neglect $\ddot\phi$ in the equation of motion for the field $\phi$, so that
\begin{equation}\label{EQMaa}
3H_0\dot\phi = - m^2 \phi\ .
\end{equation}
It is then possible to integrate the evolution equation of $\phi$,
\begin{eqnarray}
\phi(N) &=& \phi_c\,\exp(rN)\,, \nonumber\\ \label{soln}
r &\simeq& {m^2\over3H_0^2}\,,
\end{eqnarray}
where $N=H_0(t_c-t)$ is the number of $e$-folds to the phase
transition.

Quantum fluctuations of the inflaton field $\phi$ produce long
wavelength metric perturbations, ${\cal R} = H\delta\phi/\dot\phi$,
where $\delta\phi$ is the amplitude of the field fluctuation when it
crosses outside the Hubble scale. These fluctuations give rise
to a continuum spectrum of metric perturbations which can be
computed exactly in the case of hybrid inflation~\cite{GBW}
\begin{equation}\label{spectrum}
{\cal P}_{\cal R} = {C(r)^2\over r^2}\,{g^2M^2\over6\pi\lambda
M_{\rm P}^2}\,e^{-2rN}\,,
\end{equation}
where $C(r) = \Gamma[3/2-r]/2^{r}\,\Gamma[3/2]$,
and the spectral tilt can be evaluated as
\begin{equation}\label{tilt}
n-1 = {d\ln{\cal P}_{\cal R} (k)\over d\ln k} = 2r\,.
\end{equation}
Note that the tilt is always greater than one in this model.
Observations made on a wide range of scales, from COBE DMR to
CAT experiments, impose strong constraints on the amplitude and
tilt of the primordial spectrum~\cite{Charley}
\begin{eqnarray}\label{PR}
{\cal P}_{\cal R}^{1/2} &=& 5\times 10^{-5}\,(0.99\pm0.06)\,,\\
n &=& 0.91\pm0.10\,.\label{NS}
\end{eqnarray}

In the limit $m\ll H_0$, we have $r\simeq m^2/3H_0^2$, see
Eq.~(\ref{soln}), and $C(r)\simeq1$. This means that the
amplitude of the curvature perturbation spectrum should
satisfy~\cite{GBW}
\begin{equation}\label{amplitude}
{g\over\lambda\sqrt\lambda}\,{M^5\over m^2M_{\rm P}^3}
\simeq 3.5\times10^{-5}\,,
\end{equation}
while the tilt of the spectrum is bounded by~\cite{LL93}
\begin{equation}\label{tilted}
{\lambda\over\pi}\,{m^2M_{\rm P}^2\over M^4} < 0.25\,.
\end{equation}

\section{The end of inflation and the onset of the stage of
oscillations}\label{End}

As we already mentioned, in the simplest version of the hybrid
inflation scenario described above inflation ends as soon as the
scalar field $\phi$ decreases below $\phi_c = M/g$. It will be
important for us to investigate this process in a more detailed way,
because it prepares the stage for the process of reheating which we
are going to study. In particular, we should understand whether or not
the ``waterfall'' process of symmetry breaking from $\sigma = 0$ to
$|\sigma| = M/\sqrt\lambda$ can be considered as a rolling down of a
{\it homogeneous} field $\sigma$.

According to the classical equations of motion, the field $\sigma = 0$
cannot change its value because the first derivative of the effective
potential at $\sigma = 0$ vanishes. The process of spontaneous
symmetry breaking in this case occurs due to the exponential growth of
quantum fluctuations. Indeed, the field $\sigma$ has a (negative)
effective mass squared $-\mu^2(\phi) = g^2(\phi^2 - \phi_c^2)$, which
vanishes at the critical point, but becomes large and grows up to
$\mu(0) = M$ as the field $\phi$ slides towards $\phi = 0$. Quantum
fluctuations of the scalar field $\sigma$ with momentum ${\bf k}$ grow
as $\exp {w_k}t$, where $w_k =\sqrt{\mu^2-k^2}$ and $k = |{\bf k}|$.
Symmetry breaking occurs due to the growth of fluctuations with $k <
\mu$. This process produces an inhomogeneous distribution of the field
$\sigma$ with $\langle\sigma\rangle = 0$.

The resulting distribution of the scalar field $\sigma$ is relatively
homogeneous on a scale $l \sim \mu^{-1}$ or even somewhat
greater~\cite{TKKL} because the rate of exponential growth is maximal
at $k = 0$, and the fluctuations with $k > \mu$ do not grow at all.
However, in our case we have an additional complication: the effective
mass squared $-\mu^2(\phi) = g^2(\phi^2 - \phi_c^2)$ changes in time.
In order to estimate a typical scale on which the growth of the field
$\sigma$ remains relatively homogeneous we will do the following. We
will find $\mu^2$ as a function of time $\Delta t$ from the moment
when $\phi = \phi_c$. The exponential growth becomes efficient when $\mu
\sim \Delta t^{-1}$. Later, $\mu(t)$ continues growing, and all modes
$\sigma_k$ which began growing at the moment $\Delta t$, will continue
growing exponentially with approximately equal speed, and eventually
trigger a more rapid motion of the field $\phi$, see the next two
sections. This motion will preserve initial homogeneity on a scale
smaller than $\mu^{-1} \sim \Delta t$, which we are now going to
determine.

The motion of the field $\phi$ during this stage occurs due to the
negative curvature of the effective potential, $\mu^2 =
g^2(\phi_c^2-\phi^2) \approx 2g^2\phi_c |\dot \phi| \Delta t = 2 M^2
m^2\Delta t/3 H = \sqrt{2\lambda/3\pi}\, m^2M_{\rm P}\, \Delta t$.
Perturbations of the field begin to grow exponentially at $\Delta t
\sim \mu^{-1}$, which gives
  \begin{equation}\label{ccc}
\mu^3 \sim  {\sqrt\lambda }\, m^2M_{\rm P}  \ .
\end{equation}
Comparison of $\mu$ with $H \sim M^2/M_{\rm P}\sqrt{\lambda}$ shows
that $\mu \gg H$ under the ``waterfall condition'' $\lambda m M_{\rm
  P}^2 \gg M^3$ which should be satisfied in ordinary hybrid
inflation~\cite{hybrid}.  This is violated, however, in hybrid models
with two stages of inflation~\cite{Guth,GBLW}.

This means that within the time $\mu^{-1} \ll H^{-1}$ the mass squared
of the field $\sigma$ changes from being much greater than $H^2$ to
being much smaller than $- H^2$. Therefore there will be no specific
inflationary fluctuations of the field $\sigma$ with the wavelength
$H^{-1}$, but instead we will have fluctuations of this field with the
wavelength $\mu^{-1}$, which will give us the typical scale of
homogeneity $l\sim \mu^{-1} \sim (m^2M_{\rm P} {\sqrt\lambda
  })^{-1/3}$.

Now we should compare this length with other characteristic parameters
which we will consider in our study of preheating. There are two
orthogonal modes of oscillations near the minimum of the effective
potential, which we will encounter in our discussion. The first mode
corresponds to the oscillation of the field $\sigma$ at $\phi = 0$;
the corresponding frequency is given by $\bar m_\sigma = \sqrt 2 M$.
One can easily verify that $\mu \ll M$ for $\lambda M_{\rm P}^2 \gg
m^2$, which is satisfied in all hybrid inflation models. The only
possible exception known to us is the model of Refs.~\cite{Guth,GBLW}
with two different stages of inflation; we will discuss reheating in
this model separately.

Another, orthogonal mode, corresponds to the frequency $\bar m_\phi
= gM/\sqrt\lambda$, see the next section. One can show that $\mu
\ll \bar m_\phi$ if $m \ll 10^{-3} M_{\rm P} \lambda^{1/2}
g^{-3/4}$. This condition is also satisfied in most versions of the
hybrid inflation model.

The typical wavelength of perturbations produced by parametric
resonance usually is not much different from the frequency of
oscillations. Therefore our estimates indicate that in our
investigation of preheating in the usual hybrid inflation one can
treat the oscillating fields $\phi$ and $\sigma$ as homogeneous.

\section{Preheating}

Here we are interested in the behavior of the fields after the end of
inflation, as the fields oscillate around their minima, producing
particles. The qualitative behavior is very different in two opposite
limits, depending on the relation between the couplings $\lambda$ and
$g^2$, and independently of the relation among the bare masses, $m\ll
H< M$.

We will show that explosive preheating of $\phi$ and $\sigma$ particles
is not very efficient in hybrid inflation. However, we will consider
an extra scalar field $\chi$, coupled to both $\phi$ and $\sigma$,
\begin{equation}\label{couplings}
V(\chi) = {1\over2} h_1^2\,\phi^2 \chi^2 +
{1\over2} h_2^2\,\sigma^2 \chi^2\,,
\end{equation}
which will allow explosive production of $\chi$ particles in
hybrid inflation models for certain values of $h_1$ and $h_2$.

In order to study the production of particles of all fields we should
analyze the equations of motion for the quantum fluctuations of the
scalar fields. In linear perturbation theory we find

\begin{eqnarray}\label{deltaphi}
\ddot{\delta\phi}_k &+& 3H\dot{\delta\phi}_k + \Big({k^2\over a^2} +
m^2_\phi\Big)\delta\phi_k = 0 \,,\\ \label{deltasigma}
\ddot{\delta\sigma}_k &+& 3H\dot{\delta\sigma}_k + \Big({k^2\over a^2} +
m^2_\sigma\Big)\delta\sigma_k = 0 \,,\\ \label{deltachi}
\ddot\chi_k &+& 3H\dot\chi_k + \Big({k^2\over a^2} +
m^2_\chi\Big)\chi_k = 0 \,,
\end{eqnarray}
where the effective masses are
\begin{eqnarray}\label{mphi}
m^2_\phi &=& m^2 + g^2\sigma^2 + h_1^2\,\chi^2\,,\\ \label{msigma}
m^2_\sigma &=& 3\lambda\sigma^2 + g^2\phi^2 - M^2 +
h_2^2\,\chi^2\,,\\ \label{mchi}
m^2_\chi &=& h_1^2\,\phi^2 + h_2^2\,\sigma^2 \,.
\end{eqnarray}

The rate of expansion of the universe is important only in the first
few oscillations. However, it does play a role in ending the
parametric resonance regime, since modes with physical momentum $k/a$
in a resonance band will be redshifted by the expansion and fall out
of the band, shutting off particle production in that mode. It is
possible to take into account the rate of expansion by redefining the
fields $\varphi = a^{3/2} \delta\phi$, $\psi = a^{3/2} \delta \sigma$
and $X = a^{3/2} \chi$. Since the scale factor soon goes like
$a\propto t^{2/3}$, the equations of motion for these fields are the
same as Eqs.~(\ref{deltaphi})--(\ref{deltachi}) but without the
friction term proportional to $3H$. This will greatly simplify the
analysis, while still taking into account the rate of expansion.

As we mentioned in the previous section, there are two fundamental
frequencies of oscillations near the minimum of the effective
potential: $\bar m_\sigma = \sqrt 2 M$ and $\bar m_\phi =
gM/\sqrt\lambda$. Also, there are also two different scales of the
fields $\phi$ and $\sigma$: $\phi_c = M/g$, and $\sigma_0 =
M/\sqrt\lambda$.

If $\lambda \gg g^2$, one has $\sigma_0 \ll \phi_0$ and $\bar m_\sigma
\gg \bar m_\phi$. In this case oscillations of the field $\sigma$ tend
to be insignificant. This field tend to follow adiabatically the
position of the minimum of its instantaneous effective potential
depending on the value of the slowly changing field $\phi$. Most of
the energy of these two fields will be concentrated in the
oscillations of the field $\phi$. Thus the theory of reheating in this
regime should describe decay of the field $\phi$.

In the opposite limit, $\lambda \ll g^2$, one has $\sigma_0 \gg
\phi_0$ and $\bar m_\sigma \ll \bar m_\phi$. In this case the
situation is reversed.  Oscillations of the field $\phi$ tend to be
insignificant. This field tend to follow adiabatically the position of
the minimum of its instantaneous effective potential depending on the
value of the slowly changing field $\sigma$. Most of the energy of
these two fields will be concentrated in the oscillations of the field
$\sigma$, and the theory of reheating should describe decay of this
field.

The situation $\lambda \sim g^2$ is more complicated. Both fields will
oscillate with a comparable amplitude, transferring energy to each
other in a rather chaotic way.

The constants $g$ and $\lambda$ could have in principle any value, as
long as they satisfy the constraints (\ref{amplitude}) and
(\ref{tilted}) put together,
\begin{equation}
{g\over\sqrt\lambda}\,{M\over M_{\rm P}} < 5\times10^{-5}\,.
\end{equation}
For $M\sim 10^{-3} M_{\rm P}$, we have $g^2\ll\lambda$, like in the
usual hybrid inflation~\cite{hybrid}, while for $M\sim 10^{-16} M_{\rm
  P}$, we can have $g^2\gg\lambda$, like in some hybrid models with
two stages of inflation~\cite{Guth,GBLW}. Finally there is a limiting
case $g^2\sim\lambda$, which appears in the simplest models of hybrid
inflation based on supergravity, see e.g.
Refs.~\cite{CLLSW,shafi,LR,Lazar}.  Reheating in these cases is
completely different from each other. We will analyze the three
different cases in the following sections.

\subsection{Case $g^2\ll\lambda$}

\begin{figure}[t]
\centering
\hspace*{-7mm}
\leavevmode\epsfysize=6.5cm \epsfbox{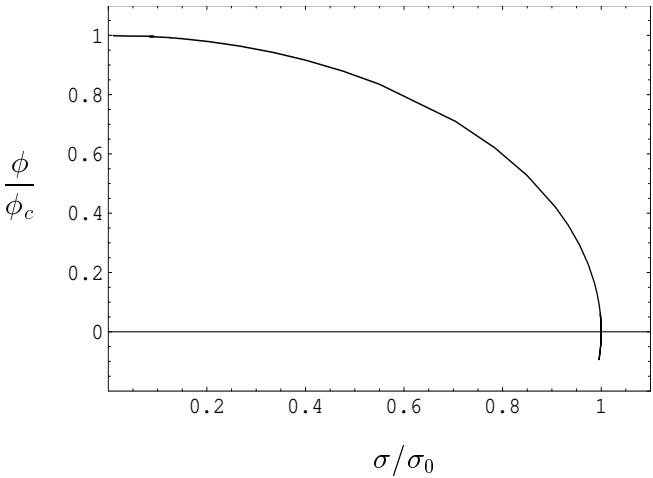}\\[2mm]
\caption[fig1]{\label{fig1} The trajectory in field space
  $(\sigma,\phi)$ after the end of inflation, along the ellipse
  $(\phi/\phi_c)^2 + (\sigma/\sigma_0)^2 = 1$. Oscillations occur
  around $\phi=0$ and $\sigma=\sigma_0$. This figure corresponds to
  $\lambda=1, g=8\times10^{-4}, M=10^{-3}\,M_{\rm P}, m=1.5\times10^{-7}\,
  M_{\rm P}$.  }
\end{figure}

\begin{figure}[t]
\centering
\hspace*{-2mm}
\leavevmode\epsfysize=6.6cm \epsfbox{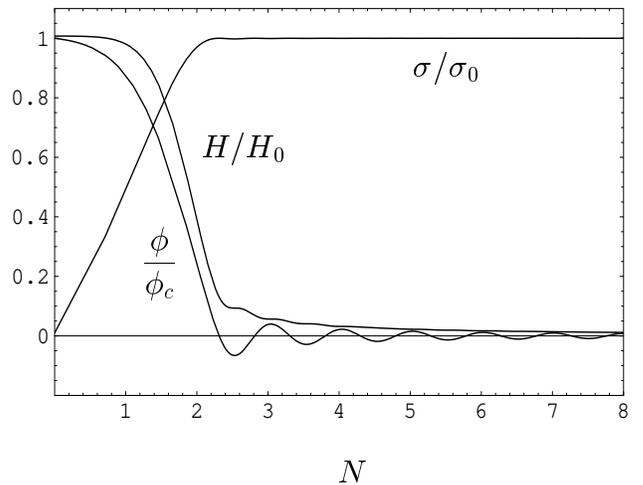}\\[2mm]
\caption[fig2]{\label{fig2} The evolution after the end of inflation
  of $H/H_0$, $\phi/\phi_c$ and $\sigma/\sigma_0$, as a function of
  the number of oscillations of the field $\phi$, $N=\bar m_\phi
  t/2\pi$.}
\end{figure}

There are two well differentiated regimes, depending on the ratio
$\bar m_\phi/H=gM/\sqrt\lambda H$. For $\bar m_\phi/H\gg1$, the amplitude of
oscillations of the inflaton field $\phi$ after inflation is large, of
order one, while for $\bar m_\phi/H\lesssim1$, the amplitude could be
very small. For both large and small ratios, the motion after the
phase transition goes along the ellipse
\begin{equation}\label{ellipse}
\lambda\sigma^2 + g^2\phi^2 = M^2 \,,
\end{equation}
in the plane $(\sigma,\phi)$, see Fig.~\ref{fig1}. As the field
$\sigma$ grows towards its minimum at $\sigma_0$, the field $\phi$
starts oscillating around zero, always along the ellipse. The main
part of the oscillation comes from the $\phi$ field (perpendicular to
the $\sigma$ direction), at $\sigma\simeq\sigma_0$,
\begin{equation}\label{Phi}
{\phi\over\phi_c} = \Phi(t) \sin \bar m_\phi t\,.
\end{equation}
In the case of Fig.~\ref{fig1}, equation (\ref{Phi}) is a good
approximation for the oscillations of the inflaton field, where
$\Phi_0 \simeq 1/15$ and $\Phi(t) \propto 1/t$, see Fig.~\ref{fig2}.
During the field's oscillations, the rate of expansion satisfies $H(t)
= 2/3t$, which corresponds to a dust like era, $a(t) \propto t^{2/3}$.

One should note that the details of the behavior of the fields $\phi$
and $\sigma$ may vary, depending on relations between different
parameters of the theory.  The results shown in Figs.~\ref{fig1}
and~\ref{fig2} correspond to $\lambda=1, g=8\times 10^{-4},
M=10^{-3}\,M_{\rm P} \sim 10^{16}$ GeV, $m=1.5\times10^{-7}\, M_{\rm
  P} \sim 2 \times 10^{12}$ GeV. In this case $\bar m_\phi$ is of the
same order as $H$ at the beginning of oscillations. This is one of the
main reasons why the amplitude of oscillations is so small.  Note that
the Hubble constant at the beginning of oscillations is approximately
ten times smaller than $H_0$. It is a very important circumstance,
because it means that even before reheating begins, the energy density
drops down by about two orders of magnitude. There are essentially no
oscillations of the field $\sigma$, and the amplitude of oscillations
of the field $\phi$ is relatively small.

However, one may consider models with different parameters, such that
$\bar m_\phi \gg H$. For example, one may consider a model with
$\lambda=1$, $g = 0.035$, $M \sim 10^{12}$ GeV and $m = 10^3$ GeV. In
this case $\bar m_\phi \sim 10^5\,H$ and the figures look somewhat
different: The amplitude of oscillations of the field $\sigma$ becomes
noticeable, but still remains very small, whereas the initial
amplitude of oscillations of the field $\phi$ becomes very large,
comparable to $\phi_c$.  This allows for a long stage of production of
$\phi$ particles in the narrow resonance regime, before the
backreaction and the expansion of the universe shifts the modes away
from the resonance, see Ref.~\cite{KLS2} and below.

It is possible to compute the production of particles of both fields
during these oscillations. Let us start with the production of
$\phi$-particles. The effective mass (\ref{mphi}) of the field $\phi$
along the trajectory (\ref{ellipse}) becomes
\begin{eqnarray}
m^2_\phi &=& m^2 + {g^2M^2\over\lambda} \Big(1 -
{\phi^2\over\phi_c^2}\Big) + h_1^2\,\chi^2\,,\nonumber\\
&\simeq& \bar m^2_\phi\,(1 - \Phi^2 \sin^2\bar m_\phi t) \,,
\end{eqnarray}
where we have neglected $m^2 \ll g^2M^2/\lambda$, see the discussion
before Eq.~(\ref{H0}), and defined $\bar m_\phi \equiv
gM/\sqrt\lambda$. Initially $\langle\chi^2\rangle = 0$, while the
field $\phi$ oscillates around zero, see Eq.~(\ref{Phi}). Substituting
into Eq.~(\ref{deltaphi}), we arrive at the Mathieu equation for the
production of $\phi$ particles,
\begin{equation}\label{Mphi}
\varphi_k'' + [A_\phi(k) -
2q_\phi \cos 2z]\,\varphi_k = 0\,,
\end{equation}
where primes denote derivatives w.r.t. $z=\bar m_\phi t$ and
\begin{eqnarray}\label{Aphi}
A_\phi(k) &=& {k^2\over a^2\bar m^2_\phi} + 1 + 2q_\phi\,,\\
q_\phi &=& {\Phi^2\over4}\,.\label{qphi}
\end{eqnarray}

To investigate parametric resonance which leads to the production of
$\phi$ particles one should find exponentially growing solutions of
the Mathieu equation (\ref{Mphi}) in an expanding universe. A detailed
explanation of this approach to preheating can be found in
Ref.~\cite{KLS2}; we will not repeat it here. We will only recall that
preheating can be especially efficient in the regime when $A_\phi(k)
\lesssim 2q_\phi$ and $q_\phi \gtrsim 1/4$, which corresponds to a
broad parametric resonance. The resonance appears not for all $k$, but
only for those $k$ belonging to the instability bands of the Mathieu
equation.  The concept of stability/instability bands is very useful
if one can neglect the role of expansion of the universe; in a more
general case one should solve Eq. (\ref{Mphi}) numerically and find
out the modes which will grow specially fast. These modes will
dominate particle production.

For a complete investigation of preheating one also needs to
investigate those effects related to backreaction of produced
particles and their rescattering~\cite{KLS2,GKLS,Tkachev}. We will
make some comments on this issue later, but we will mainly concentrate
on the first stages of preheating, where the corresponding effects can
be neglected. This approach will allow us to identify those versions
of hybrid inflation where preheating may be efficient.

For the model described by Figs.~\ref{fig1},~\ref{fig2} (with
$\lambda=1, g = 8\times 10^{-4}, M=10^{-3}\,M_{\rm P},
m=1.5\times10^{-7}\, M_{\rm P}$) the initial value of $q$ is extremely
small, $q_\phi = \Phi_0^2/4 \simeq 10^{-3}\ll~1$. For $k\simeq0$, we
are in the narrow region of the first resonance band, $A=1$.  In this
case there are solutions of Eq.~(\ref{Mphi}) that correspond to
exponential instabilities with occupation numbers of quantum
fluctuations given by $n_k(t)\propto \exp(2\mu_k z)$, which can be
interpreted as particle production~\cite{KLS1}. However, the effective
parameter $2\mu_k \sim q$ will be extremely small and particle
production will not be very efficient.

It is possible to evaluate the time it takes for a particular mode $k$
to cross out of the resonance band, $\Delta t\sim\mu_k H^{-1}$, see
Ref.~\cite{KLS1}.  The occupation number then becomes
$n_k\sim\exp\,(2\mu_k^2\bar m_\phi/H)$. Explosive production shuts down
when $\Phi^4 < 32H/\bar m_\phi$. In our model, this is true from the
very beginning, and so we are confident that very little production of
$\phi$-particles will occur.

In the models where the initial amplitude of the field $\phi$ is large
(of the same order as $\phi_c$), production of $\phi$ particles can be
much more efficient.  For example, in the model with $\lambda=1$, $g =
0.035$, $M \sim 10^{12}$ GeV and $m = 10^3$ GeV one has case $\bar
m_\phi \sim 10^5\,H$, which means that the field can make about $10^4$
oscillations before expansion of the universe becomes important. The
initial value of the parameter $q$ is about $1/4$, see (\ref{qphi}).
This corresponds to the narrow resonance regime in the second
instability band of the Mathieu equation.

The resonance in the second instability band at small $q$ typically is
very narrow and inefficient. Moreover, one should check carefully
whether this band actually exists, because Eq.~(\ref{Mphi}) is correct
only when one neglects oscillations of the field $\sigma$. Meanwhile
these oscillations are not entirely negligible, so they may affect the
narrow resonance in the second instability band. And indeed in our
numerical investigation of the resonance in the coupled system of the
fields $\phi$ and $\sigma$ we did not find any evidence of $\phi$
particle production. Even if this process happens, it is not really
reheating; parametric resonance breaks the coherently oscillating
field $\phi$ into a collection of $\phi$ particles. A subsequent
perturbative decay of $\phi$ particles is necessary.

Let us consider now the production of $\sigma$-particles. The
effective mass (\ref{msigma}) along the trajectory (\ref{ellipse})
becomes
\begin{eqnarray}\label{barms}
m^2_\sigma &=& 2M^2 \,\Big(1 - {\phi^2\over\phi_c^2}\Big) +
h_2^2\,\chi^2\,,\nonumber \\
&\simeq& \bar m^2_\sigma \,(1 - \Phi^2 \sin^2\bar m_\phi t)  \,,
\end{eqnarray}
where initially $\langle\chi^2\rangle = 0$ and $\bar m_\sigma \equiv
\sqrt2\,M$. Substituting back into Eq.~(\ref{deltasigma}), we arrive
at the Mathieu equation for the production of $\sigma$ particles,
\begin{equation}\label{Msigma}
\psi_k'' + \left[A_\sigma(k) -
2q_\sigma \cos 2z\right]\psi_k = 0\,,
\end{equation}
where $z=\bar m_\phi t$ and
\begin{eqnarray}\label{Aqsigma}
A_\sigma(k) &=& {k^2\over a^2\bar m^2_\phi} +
{\bar m^2_\sigma\over\bar m^2_\phi} + 2q_\sigma\,,\\
q_\sigma &=& {\bar m^2_\sigma\over\bar m^2_\phi} \,{\Phi^2\over4}\,.
\end{eqnarray}
The ratio $\bar m^2_\sigma/ \bar m^2_\phi = 2\lambda/g^2$, while
$q\simeq(\lambda/g^2)\,\Phi^2/2$.  Since $\Phi\ll2\sqrt{2\lambda}/g$
in our case, we have that $A-2q\gg\sqrt q$, i.e. far above the
resonance band, and therefore we expect no explosive production of
$\sigma$-particles from the oscillations of the field $\phi$.

Moreover, $\sigma$ particles can neither be produced at the next
(perturbative) stage of the decay of the field $\phi$. Indeed, the
effective mass of $\phi$-particles, $\bar m_\phi \sim
gM/\sqrt\lambda$, in the case $g^2 \ll \lambda$, is much smaller than
the mass of the $\sigma$ particles, $\sqrt 2 M$, so the decay of
$\phi$ particles is kinematically impossible. Thus, in this regime the
field $\phi$ cannot transfer its energy to $\sigma$ particles.

It is possible, however, to parametrically amplify fluctuations of
another scalar (or vector) field $\chi$, which interacts with the
fields $\phi$ and $\sigma$ in accordance with Eq.~(\ref{couplings}).
Since oscillations will occur around $\phi=0$, while
$\sigma\simeq\sigma_0$, in order to have particle production in this
field we require that the induced mass from the symmetry breaking
field $\sigma$ be much smaller than the corresponding oscillations
from the inflaton field $\phi$, $h_1^2\phi^2 \gg h_2^2\sigma_0^2$,
which at the beginning of preheating corresponds to
\begin{equation}\label{hcond}
\lambda h_1^2 \gg g^2 h_2^2\,.
\end{equation}
It is always possible to find couplings $h_1, h_2$ that satisfy this
constraint. Eventually, when the amplitude of oscillations of the
$\phi$ field becomes
\begin{equation}\label{phicond}
\Phi(t) < {h_2\,g\over h_1 \sqrt\lambda}\,,
\end{equation}
the explosive production of $\chi$-particles will end. This may happen
before or after backreaction sets in, depending on the parameters of
the model.

\begin{figure}[t]
\centering
\hspace*{-3.5mm}
\leavevmode\epsfysize=5.9cm \epsfbox{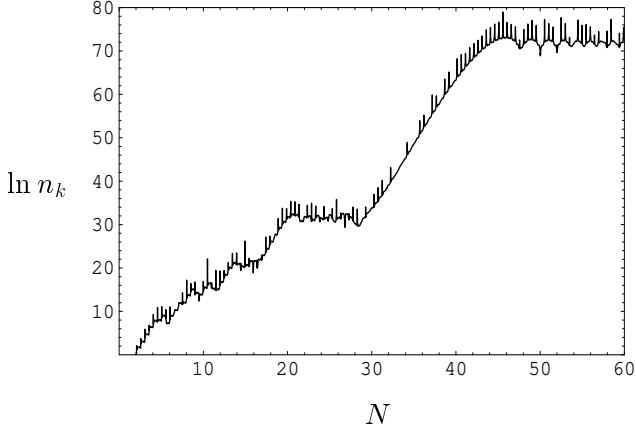}\\[2mm]
\caption[fig3]{\label{fig3} The exponential growth of the occupation
  number $n_k$ of $\chi$-particles, for $k=\bar m_\phi$, as a function
  of the number of oscillations of the field $\phi$, $N=\bar m_\phi
  t/2\pi$. One can distinguish here the broad stochastic resonance
  regime. The typical value of the growth parameter is $\mu \simeq
  0.13$ during the last stages of preheating.}
\end{figure}

The corresponding Mathieu equation for the production of $\chi$-particles
is given by
\begin{equation}\label{Mchi}
X_k'' + [A(k) - 2q \cos 2z]\,X_k = 0\,,
\end{equation}
where
\begin{eqnarray}\label{Achi}
A(k) &=& {k^2\over a^2\bar m^2_\phi} + 2q \,,\\ \label{qchi}
q &=& {h_1^2 M^2\over g^2\bar m^2_\phi}\,{\Phi^2\over4}\,.
\end{eqnarray}
Initially, $q_0 = (\lambda h_1^2/g^4)\,\Phi_0^2/4 \simeq (\lambda
h_1^2/g^4)\,10^{-3}$, where we have assumed $h_2$ satisfies
Eq.~(\ref{hcond}) and can be neglected.  Since $\lambda\gg g^2$, it is
possible to find natural values of the parameter $h_1$ that ensure
$q_0 \gg 1$. We are therefore in the broad parametric resonance region
and explosive production of $\chi$-particles will occur. In this case
there are solutions of Eq.~(\ref{Mchi}) that correspond to exponential
instabilities with occupation numbers of quantum fluctuations given by
\begin{eqnarray}
n_k(t) &=& {\omega_k\over2}\Big({|\dot X_k|^2\over\omega_k^2} +
|X_k|^2\Big) - {1\over2} \nonumber \\ \label{occnum}
&\simeq& {1\over2} \exp(2\mu_k z)\,,
\end{eqnarray}
which can be interpreted as particle production~\cite{KLS1},
where the frequency $\omega_k$ is given by
\begin{equation}\label{omega}
\omega_k^2 = {k^2\over a^2} + h_1^2\,\phi^2 \,.
\end{equation}
We have shown the exponential growth of the occupation number $n_k$ of
$\chi$-particles in Fig.~\ref{fig3}a, together with the effective
growth parameter $\mu_k$ in the broad resonance region, see
Eq.~(\ref{occnum}) and Fig.~\ref{fig3}b. The model parameters used are
$\lambda=1, h_1=g=8\times 10^{-3}, M=10^{-3}\,M_{\rm P},
m=1.5\times10^{-7}\, M_{\rm P}$, for the typical momentum $k=\bar
m_\phi$. In this case, the typical growth parameter is relatively
large, $\mu\simeq0.13$, and particle production will be very
efficient. Furthermore, for a large range of couplings we will enter
the region of stochastic resonance, see Ref.~\cite{KLS2}.

The present case is analogous to the case found in~\cite{KLS2} for
a massive inflaton field, where we substitute $m \to \bar m_\phi =
gM/\sqrt\lambda$ and $g \to h_1$. All features found in that case will
be present here, for a certain range of parameters. For instance,
backreaction will set in at the time $t_1$ given by~\cite{KLS2}
\begin{equation}\label{N1}
N_1 = {\bar m_\phi t_1\over2\pi} =
{5\over8\pi\mu}\,\ln{15\over h_1} \simeq 15\,,
\end{equation}
when the expectation value of created particles satisfies
\begin{eqnarray}\label{backreac}
\langle\chi^2\rangle_1 \simeq {\bar m_\phi^2\over h_1^2} =
{g^2 M^2\over\lambda\,h_1^2}\,,
\end{eqnarray}
and the oscillating field $\phi$ acquires a large effective mass.
At that time the parameter $q$ becomes
\begin{equation}\label{q1}
q_1^{1/2} = {q_0^{1/2}\over4N_1} \simeq {h_1\sqrt\lambda\over4g^2}\,
10^{-4}\,.
\end{equation}
The efficiency of preheating at this stage is determined by the
fraction of energy density in kinetic energy of $\chi$ particles,
$\langle(\nabla\chi)^2\rangle_1 = \bar m_\phi^2\phi_c^2\Phi^2_1\,
q_1^{-1/2}$, see Ref.~\cite{KLS2}, which could be very small compared
with the energy density remaining in the oscillations of the inflaton
field, if $q_1\gg~1$.

Even after backreaction there is still production of $\chi$-particles
in the broad resonance regime until the parameter $q$ falls below
$1/4$, see Ref.~\cite{KLS2}. At that time,
\begin{eqnarray}\label{narrow}
\langle\chi^2\rangle_2 \simeq \phi_c^2\,\Phi^2(t_2) =
{4g^2M^2\over\lambda\,h_1^2}\,q_1^{1/2}\,,
\end{eqnarray}
where we have used the relation $\Phi_2 \simeq \Phi_1\,q_1^{-1/4}$
between the amplitude of oscillations at both times, see~\cite{KLS2}.
At this time the energy density is equally distributed between the
kinetic energy of the $\chi$ particles and their interaction energy
with the field $\phi$. In that case, $\langle(\nabla\chi)^2\rangle_2 =
\bar m_\phi^2\phi_c^2\Phi^2_2\,q_1^{1/2} = \bar
m_\phi^2\phi_c^2\Phi^2_1$, see Ref.~\cite{KLS2}. This result gets
modified when including rescattering of $\chi$ particles with
particles of the field $\phi$. The end of preheating occurs slightly
earlier, when $\Phi_r \simeq 2.5 \Phi_1\,q_1^{-1/4}$, and the energy
density is predominantly in the interaction energy between $\chi$
particles and the field $\phi$, $\rho_\chi \sim 10^{-2} g^2 \phi_c^4
\Phi_r^4$.

A relevant question in the presence of preheating is whether
non-thermal effects due to large occupation numbers of particles can
restore the symmetry in the $\sigma$ field, see Ref.~\cite{KLS2}.
For this to occur, we need that the effective mass in the false vacuum
be positive at the end of preheating,
\begin{eqnarray}\label{sr}
V''_\sigma(0) &=& - M^2 + h_2^2\,\langle\chi^2\rangle_2\\ \label{effmass}
&=& M^2 \Big(- 1 + {4g^2 h_2^2\over\lambda\,h_1^2}\,q_1^{1/2}\Big) >0
\end{eqnarray}
We see that the condition for symmetry restoration in this model
requires
\begin{equation}\label{srcond}
q_1^{1/2} = {q_0^{1/2}\over4N_1} >
{\lambda\,h_1^2\over4g^2 h_2^2} \gg 1\,,
\end{equation}
see Eq.~(\ref{hcond}). Unless the initial value $q_0$ is very large
indeed, it is very difficult to satisfy this condition. There is
however no fundamental reason why this should not happen, and, as a
consequence, non-thermal symmetry restoration could in principle be
possible in this model for certain relations between coupling constants. If
this happens, the description of preheating given above will need to be
modified.

Let us now study the opposite limit, $\lambda\ll g^2$, and see whether
there is parametric resonance production of particles in that case.

\subsection{\label{small}Case $\lambda\ll g^2$}

For $\lambda\ll g^2$, the most important mode of oscillations is the
oscillation of the field $\sigma$ near the minimum of the effective
potential at $\sigma_0 = M/\sqrt\lambda$. The initial stages of
oscillations can be very complicated. Depending on the relations
between the parameters of the theory, the first motion of the fields
$(\phi,\sigma)$ may occur either in the $\sigma$-direction, as in the
standard hybrid inflation scenario \cite{hybrid}, or in $\phi$
direction, as in Ref. \cite{Guth,GBLW}. However, in both cases the
oscillations of the field $\phi$ eventually become small, and all
energy becomes concentrated in the oscillations of the field $\sigma$,
see Fig. ~4.

\begin{figure}[t]
\centering
\hspace*{-4mm}
\leavevmode\epsfysize=6.6cm \epsfbox{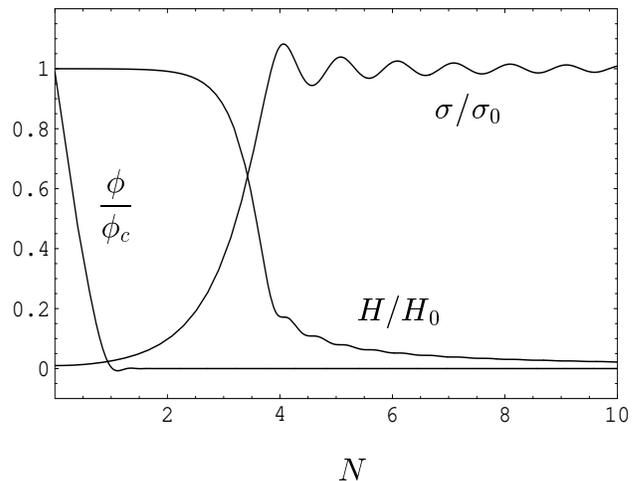}\\
\caption[fig4]{\label{fig4} The evolution after the end of inflation
  of $H/H_0$, $\phi/\phi_c$ and $\sigma/\sigma_0$, as a function of
  the number of oscillations of the field $\sigma$, $N=\bar m_\sigma
  t/2\pi$. This figure corresponds to $\lambda=2\times10^{-16}, g
  =1.4\times10^{-5}, M=m=10^{-8}\, M_{\rm P}$.  }
\end{figure}

Thus, just as in the case $\lambda \gg g^2$, the motion of the fields
eventually becomes essentially one-dimensional, but now it occurs in
the $\sigma$ direction. The corresponding process is very similar to
the one which occurs in new inflation. The oscillations of the field
$\sigma$ can be represented in the following way:
\begin{equation}\label{Sigma}
{\sigma\over\sigma_0} = 1 + \Sigma(t)\,\sin \bar m_\sigma t\,.
\end{equation}
This is a good approximation for the oscillations of the $\sigma$
field with a small amplitude $\Sigma_0 \ll 1$, $\Sigma(t)\propto 1/t$,
see Fig.~\ref{fig4}.

It is then possible to compute the production of particles of both
fields during the $\sigma$-oscillations.  Let us start with the
production of $\sigma$-particles. Before the symmetry breaking field
starts oscillating, the field $\phi$ has negligible amplitude, so we
will assume $\phi\simeq 0$. The effective mass (\ref{msigma}) of the
field $\sigma$ then becomes
\begin{eqnarray}
m^2_\sigma &=& 2M^2 + 3\lambda \,(\sigma^2-\sigma_0^2) +
h_2^2\,\langle\chi^2\rangle\,, \nonumber \\ \label{tildemsigma}
&\simeq& \bar m^2_\sigma(1 + 3\Sigma\,\sin\bar m_\sigma t) \,,
\end{eqnarray}
where $\langle\chi^2\rangle = 0$ initially, and $\bar m^2_\sigma
\equiv 2M^2$. Substituting back into Eq.~(\ref{deltasigma}), we arrive
at the Mathieu equation for the production of $\sigma$ particles,
\begin{equation}\label{Msigma2}
\psi_k'' + \left[A_\sigma(k) +
2q_\sigma \sin 2z\right]\,\psi_k = 0\,,
\end{equation}
where $z=\bar m_\sigma t/2$ and\footnote{
We have neglected the $\Sigma^2$ and the $\Phi^2$ terms,
which would have contributed as higher harmonics and soon be completely
irrelevant.}
\begin{eqnarray}\label{Asigma2}
A_\sigma(k) &=& {2k^2\over a^2 M^2} + 4 \,,\\
\label{qsigma2}
q_\sigma &=&  6\,\Sigma\,.
\end{eqnarray}
Initially in our model we have $\Sigma \simeq 1/12$,
$\langle\chi^2\rangle = 0$, so $q_\sigma = 6\,\Sigma \simeq 0.5$, see
Fig.~\ref{fig4}. For $k\simeq0$, we are in the narrow region of the
second resonance band, $A=4$. There will some production of
$\sigma$-particles but very soon the modes will redshift away and
production will shut down.

The effective mass of $\phi$ particles
(\ref{mphi}) for small $\phi$  and $m \ll \bar m_\phi$ becomes
\begin{eqnarray}
m^2_\phi &=& m^2 + {g^2M^2\over\lambda} +
g^2\,(\sigma^2-\sigma_0^2) + h_1^2\,\langle\chi^2\rangle\,,\nonumber\\
\label{tildemphi}
&\simeq& \bar m^2_\phi (1 + 2 \Sigma\,\sin\bar m_\sigma t) \,,
\end{eqnarray}
where $\langle\chi^2\rangle=0$ initially, and $\bar m_\phi =
gM/\sqrt\lambda$. This corresponds to the Mathieu parameters
\begin{eqnarray}\label{Aphi2}
A_\phi(k) &=& {2k^2\over a^2 M^2} + {2 g^2\over \lambda} \,,\\
q_\phi &=&  {2 g^2\over \lambda}\,\Sigma\,.\label{qphi2}
\end{eqnarray}
Therefore, for $\Sigma \ll 1$ one has $q_\phi \ll A_\phi(k)$. In this
regime the resonance is very narrow, and one does not expect a large
production of $\phi$ particles.

This conclusion remains valid for the perturbative decay of the
oscillating scalar field $\sigma$ as well.  Indeed, the effective mass
of the field $\phi$ particles $\bar m_\phi \sim gM/\sqrt\lambda$ in
the case $g^2 \gg \lambda$ is much greater than the mass of the
$\sigma$ particles $\sqrt 2 M$, so the decay of $\sigma$ particles to
$\phi$ particles is kinematically impossible.  Thus, in this regime
the field $\phi$ cannot transfer its energy to $\sigma$ particles.
This is very similar to what happens in the case $g^2 \ll \lambda$,
where the energy is concentrated in the oscillating field $\phi$,
which (for $g^2 \ll \lambda$) cannot decay to $\sigma$ particles, see
the previous section.

Parametric resonance with production of $\chi$ particles in principle
could be possible. In the case where masses of particles near the
minimum of the effective potential are not much greater than $H$, one
can ignore the small $\phi$ oscillations. Then the $\chi$ particle
production is described by the Mathieu equation
\begin{equation}\label{Mux}
\ddot X_k + [{k^2\over a^2} + h_2^2\sigma_0^2 +2h_2^2\sigma_0^2
\Sigma(t)  \sin \bar m_\sigma  t ]\,X_k = 0 \ ,
\end{equation}
see~\cite{KLS2}. A change of variables $\bar m_\sigma t = 2z -\pi /2$
reduces Eq. (\ref{Mux}) to the Mathieu equation
\begin{equation}\label{M1ux}
X_k'' + [A(k) - 2q \cos 2z ] X_k = 0 \ .
\end{equation}
Here $A(k) = 2k^2/a^2M^2 + 2h_2^2/\lambda$, $q =
(2h_2^2/\lambda)\,\Sigma$, $z = \bar m_\sigma t/2$, and primes denote
differentiation with respect to~$z$.

Note that $q < A(k) \Sigma$. In this case the Mathieu plot of
stability/instability bands shows that as soon as the amplitude of
oscillations of the field $\sigma$ becomes much smaller than
$\sigma_0$, one has $\Sigma \ll 1$, which implies that $q \ll A(k)$.
In this regime the resonance is narrow, and preheating is not very
efficient~\cite{KLS1}. In our case $\Sigma \ll 1$ from the very
beginning of the period of oscillations, and $\chi$ particle
production is inefficient.

It is possible in principle to conceive a scalar field
$\chi$ which couples not to $\sigma$ but to the difference $\sigma -
\sigma_0$,
\begin{equation}\label{mchi2}
m_\chi^2 = h_2^2\,(\sigma-\sigma_0)^2\,
\end{equation}
In such models production of $\chi$-particle can be very efficient.
Let us analyze this case is more detail here.

\begin{figure}[t]
\centering
\hspace*{-4mm}
\leavevmode\epsfysize=6.1cm \epsfbox{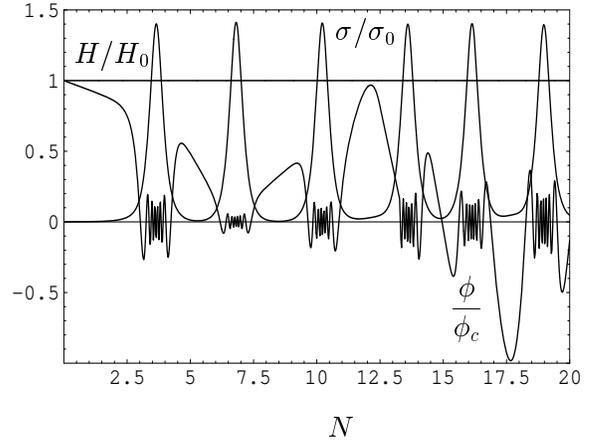}\\[2mm]
\caption[fig5]{\label{fig5} The evolution after the end of inflation
  of $H/H_0$, $\phi/\phi_c$ and $\sigma/\sigma_0$, as a function of
  $N=\bar m_\sigma t/2\pi$, in the case $M \gg H$. Note that the
  number of oscillations of the field $\sigma$ is about $N/3$, while
  the amplitude of oscillations of both fields is large. This figure
  corresponds to $\lambda=10^{-2}, g=1, m=10^3\,{\rm GeV},
  m=1.3\times10^{11}\,{\rm GeV}$.  }
\end{figure}

Substituting (\ref{mchi2}) into Eq.~(\ref{deltachi}), we arrive
at the Mathieu equation for the production of $\chi$ particles,
(\ref{Mchi}), with $z = \bar m_\sigma t$ and
\begin{eqnarray}\label{Achi2}
A(k) &=& {k^2\over a^2\bar m_\sigma^2} + 2q \,,\\ \label{qchi2}
q &=& {h_2^2 M^2\over\lambda\bar m_\sigma^2}\,{\Sigma^2\over4}\,.
\end{eqnarray}
Initially, $\langle\chi^2\rangle = 0$, so $q_0 =
(h_2^2/\lambda)(\Sigma^2/8) \simeq (h_2^2/\lambda)\,10^{-3}$, see
Fig.~\ref{fig4}. Since $\lambda \ll g^2$ in this case, we can easily
get $q_0\gg1$. This ensures explosive $\chi$-particle production with
a large effective growth parameter $\mu_k$. As an example, we
considered a model with $\lambda=2\times10^{-16},
h=10\,g=1.4\times10^{-5}, M=m=10^{-8}\, M_{\rm P}$. We have found that
the typical growth parameter is rather large, $\mu\simeq0.13$, and
particle production is very efficient.

\begin{figure}[t]
\centering
\hspace*{-4mm}
\leavevmode\epsfysize=5.9cm \epsfbox{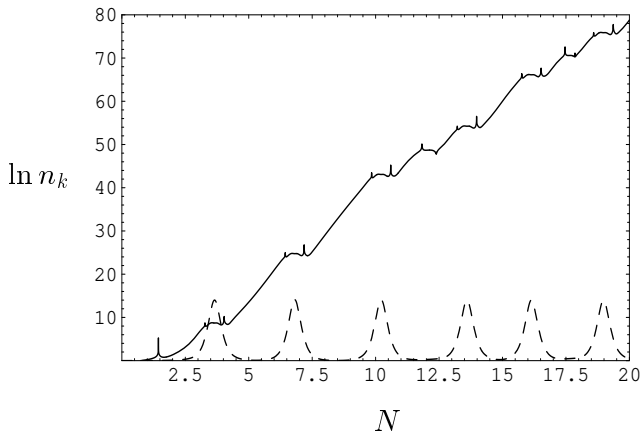}\\[3mm]
\hspace*{-4mm}
\leavevmode\epsfysize=5.9cm \epsfbox{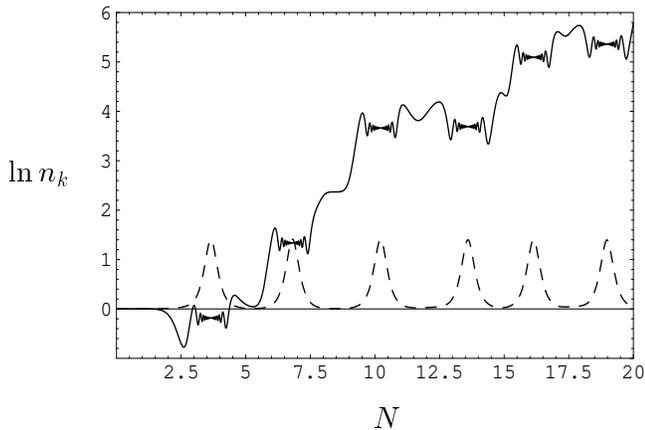}\\[2mm]
\caption[fig5p]{\label{fig5p} The top panel shows the exponential
growth of the occupation number $n_k$ of $\sigma$ particles, as a
function of $N=\bar m_\sigma t/2\pi$, for $k=0.43M$ and the parameters
of Fig.~\ref{fig5}. It acquires a typical growth parameter $\mu_k
\simeq 0.3$ during the last stages of preheating. The lower panel
shows the occupation number $n_k$ of $\phi$ particles, for
$k=0.2M$. The growth parameter is an order of magnitude smaller,
$\mu_k \simeq 0.023$, in this case. The dashed line shows the
oscillations of the field $\sigma$. The upper panel shows
$10\sigma/\sigma_0$, while the lower panel shows $\sigma/\sigma_0$. 
The number of the $\sigma$ field oscillations differs from 
$N=\bar m_\sigma t/2\pi$ because the oscillations are not harmonic.}
\end{figure}

Note however that it is rather difficult (though not impossible) to
invent a realistic model where $\chi$ particles have masses given by
Eq.~(\ref{mchi2}). We have found that it is possible in the context of
supersymmetric models, but it requires explicit fine tuning of the
superpotential of the theory. Thus, one may conclude that in a general
class of hybrid inflation models where $\chi$ particles acquire a mass
$h_2\sigma_0$ after spontaneous symmetry breaking, preheating for
$\lambda \ll g^2$ is inefficient.  This question should be reexamined
in the theories with flat directions, and in the theories where $\chi$
particles have small mass due to an accidental cancellation of two
large contribution. Such a situation may not be entirely unrealistic;
remember that the Higgs doublet of the standard model has an extremely
small mass as compared to the GUT scale.

\begin{figure}[t]
\centering
\hspace*{-4mm}
\leavevmode\epsfysize=6.0cm \epsfbox{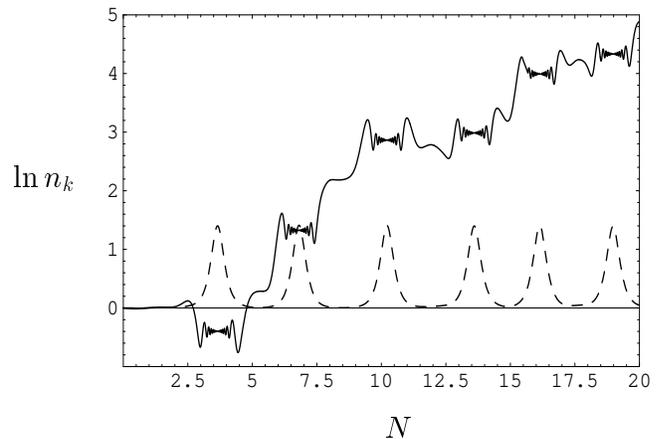}\\[2mm]
\hspace*{-4mm}
\leavevmode\epsfysize=6.0cm \epsfbox{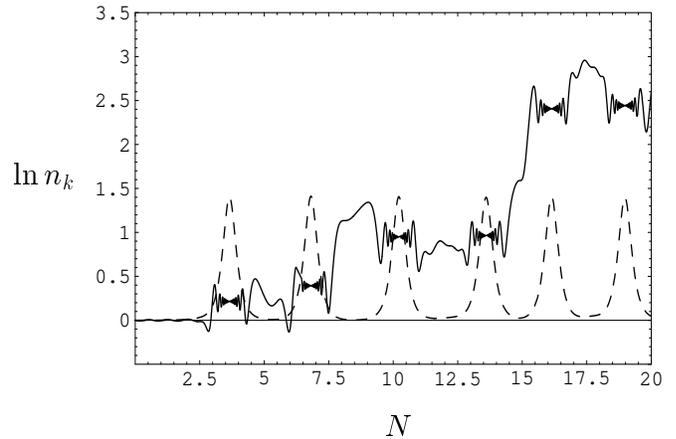}\\[2mm]
\caption[fig5pp]{\label{fig5pp} The top panel shows the occupation
  number $n_k$ of $\chi$ particles as a function of $N=\bar m_\sigma
  t/2\pi$, for $k=0.2M,\ h_1=0.3,\ h_2=1$ and the parameters of
  Fig.~\ref{fig5}. It acquires a typical growth parameter $\mu \simeq
  0.03$ during the last stages of preheating.  The lower panel shows
  the same as above but for $h_1=h_2=1$. The growth parameter is
  somewhat smaller, $\mu \simeq 0.013$, in this case. The dashed line
  shows $\sigma/\sigma_0$.}
\end{figure}

In the version of the hybrid inflation scenario which we discussed
until now the amplitude of oscillations of the field $\sigma$ was
relatively small.  However, our conclusions concerning parametric
resonance for $\lambda \ll g^2$ can change considerably if one finds
the models with $M \gg H$, which induce a quick end of inflation and a
large amplitude of oscillations of the field $\sigma$ after inflation.
For example, one may consider a model with $g = 1$, $\lambda =
10^{-2}$, $m = 10^3$ GeV, $M = 1.3 \times 10^{11}$ GeV.  This gives $H
\sim 2\times 10^4$ GeV, $\bar m_\phi = 5\times 10^7 H$ and $\bar
m_\sigma = 8\times 10^6 H$. In such a model the amplitude of the
oscillations of the field $\sigma$ is very large, $\Sigma \sim 1$, see
Fig.~\ref{fig5}. Moreover, this amplitude will not be damped by the
expansion of the universe during $10^{6}$ oscillations of this field.
As a result, in this model there will be a very efficient production
of $\sigma$, $\phi$ and $\chi$ particles in a broad resonance regime,
see Figs.~\ref{fig5p} and~\ref{fig5pp}.

These figures illustrate the growth of fluctuation neglecting
backreaction. In fact, preheating in this model is so efficient that
it completes within few oscillations. Indeed, let us estimate how
strong should be the growth of $n_k$ in order to transfer all energy
of the oscillating scalar field $\sigma$ to $\sigma$ particles (this
is the leading process). The initial energy density of the field
$\sigma$ is $M^4/4\lambda$. Each produced particle carries the
energy $\sim \bar m_\sigma \sim \sqrt 2 M$.  This implies that one
should have $n_k \sim \lambda^{-1}$ in order to transfer all energy of
oscillations into $\sigma$ particles.  The total duration of the
process in terms of the number of oscillation N follows from the
relation $\exp(4\pi \mu N ) \sim \lambda^{-1}$, which gives $N \sim
(4\pi\mu)^{-1} \ln \lambda^{-1}$. For $\mu \sim 0.3$ and $\lambda \sim
10^{-2}$ one has $N = O(1)$, i.e. the whole process finishes within
one or two oscillations; the   part of Figs.~\ref{fig5p} and~\ref{fig5pp}
 at $N\gg 1$ is redundant.

The unusual efficiency of the process of $\sigma$ particle production
can be explained by the relatively large value of the coupling
constant $\lambda$ and by the tachyonic instability of the field at
the first stages of its rolling down. A detailed theory of this
process will be considered elsewhere~\cite{TKKL}. A simple
interpretation of our results is as follows.

The process of falling down of the field $\sigma$ in the regime
$\lambda \ll g^2$ occurs almost independently of the field $\phi$. The
field $\phi$ triggers the process, but then the field $\sigma$ moves
essentially as a field falling down from the point $\sigma = 0$ in the
theory with the effective potential ${\lambda\over
4}(\sigma^2-\sigma_0)^2$. As we discussed in Section \ref{End},
neglecting expansion of the universe, fluctuations of the field
$\sigma$ in the beginning of the process grow as
$\exp\sqrt{k^2-M^2}t$. Initially, there was no homogeneous classical
field $\sigma$, the whole process of spontaneous symmetry breaking
occurs due to the growth of quantum fluctuations. The value of the
homogeneous mode $\langle \sigma\rangle$ vanishes all the time;
spontaneous symmetry breaking occurs when $\sqrt{\langle
\sigma^2\rangle}$ created by the instability approaches $\sigma_0$. As
a result, already after the quarter of the first oscillation,
$\sqrt{\langle \sigma^2\rangle}$ approaches $\sigma_0$, all energy
will be concentrated not in a homogeneous field $\sigma$ but in its
fluctuations with a wavelength $l \sim 2\pi M^{-1} \ln^{1/2}
{4\pi^2\over \lambda}$, which is somewhat greater than $M^{-1}$
\cite{TKKL}. In this sense one may even say that preheating ends
within one quarter of an oscillation. This, however, would be not
quite correct because after rolling there will be a stage of resonant
production of particles with momenta comparable to $M$.

We should emphasize, however, that here again
we deal not with the decay of the field $\sigma$ to other particles,
but with a rapid process of transformation of the energy of the   classical
field
$\sigma$ to the energy of $\sigma$ particles. The subsequent decay of these
particles occurs in accordance with the elementary theory of reheating
\cite{reheat}.

Let us now briefly discuss the case when the two couplings $g^2$ and
$\lambda$ are of similar magnitude.

\subsection{Case $g^2\simeq\lambda$}

This case corresponds to the simplest supergravity hybrid inflation
model~\cite{CLLSW,shafi,LR}, based on a superpotential of the type
\begin{equation}\label{sugra}
W = S\,(\kappa\,\bar\phi\phi - \mu^2)
\end{equation}
which induces a scalar potential with self-coupling $\lambda =
\kappa^2$ for the triggering field $\phi$, and couples the two fields
with strength $g^2 = \kappa^2$, see Eq.~(\ref{hybrid}). Thus, we have
the same coupling, $\lambda = g^2$, as one should expect from
supersymmetry universality.\footnote{This is not a general rule. For
  example, it is not necessarily true in D-term hybrid
  inflation~\cite{D}, where the couplings $g^2$ and $\lambda$ come
  from the F and the D terms respectively.  }

\begin{figure}[t]
\centering
\hspace*{-4mm}
\leavevmode\epsfysize=6.5cm \epsfbox{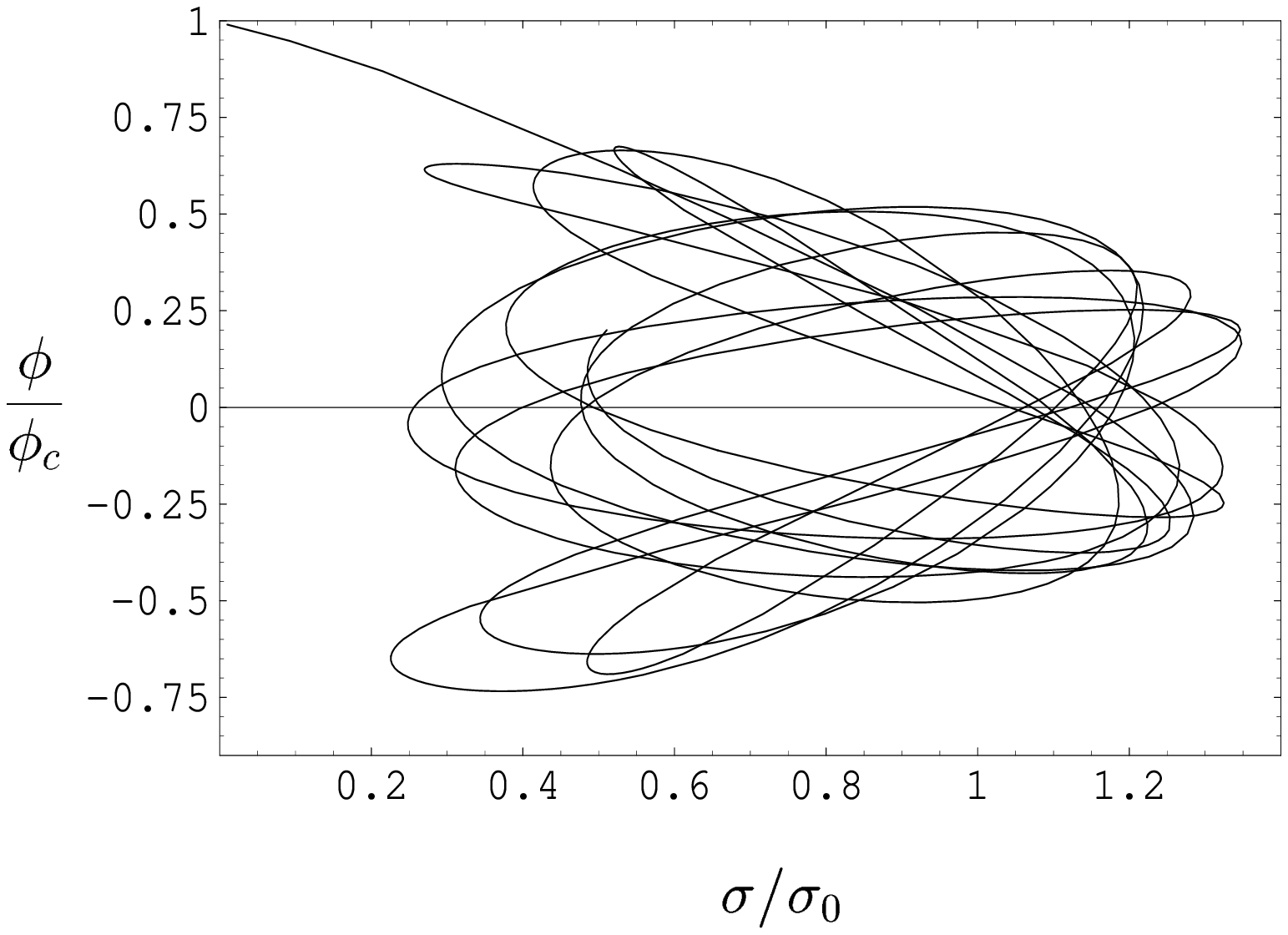}\\[3mm]
\hspace*{-4mm}
\leavevmode\epsfysize=6.1cm \epsfbox{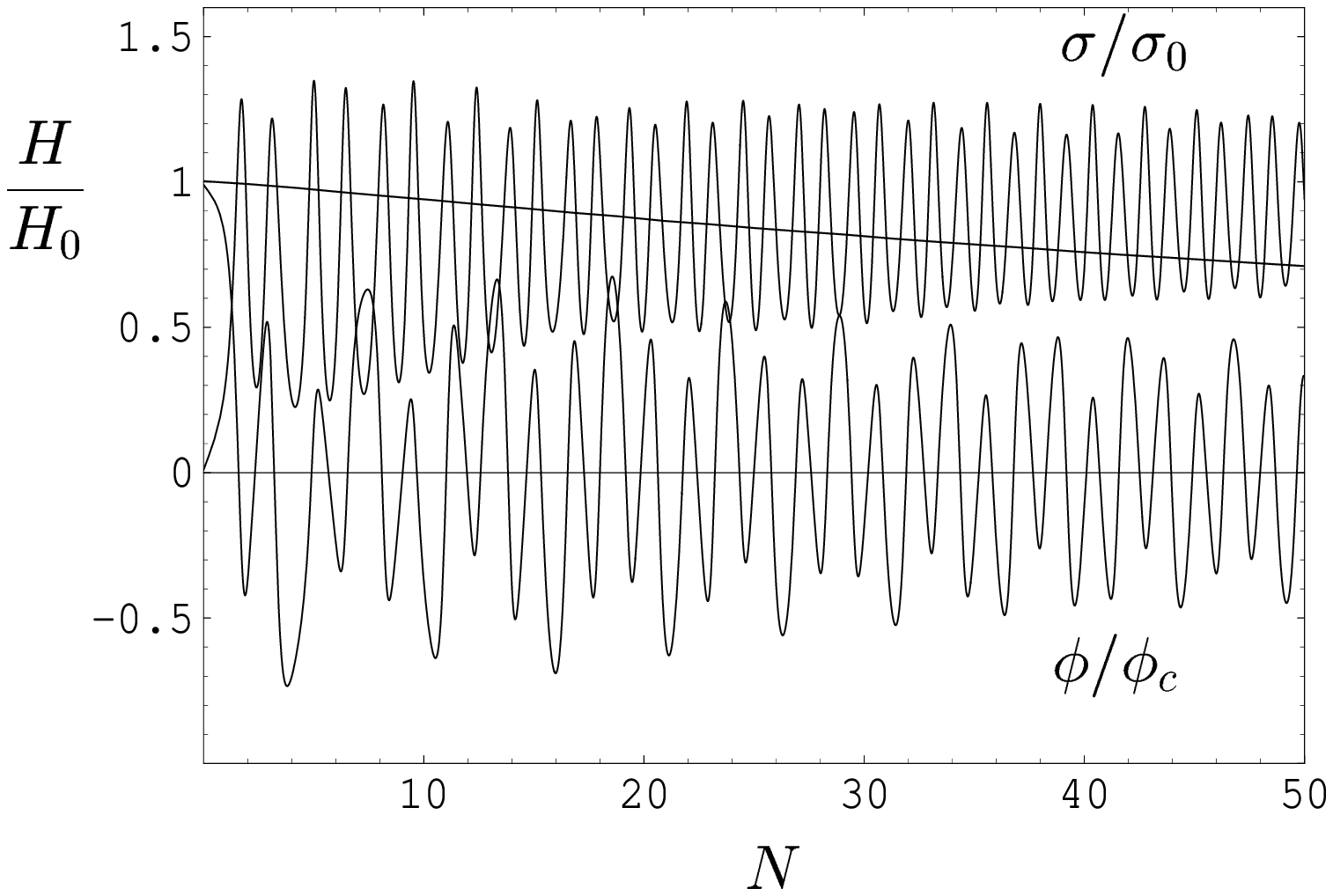}\\[2mm]
\caption[fig6]{\label{fig6} The top panel shows the motion of the
  two scalar fields in the plane ($\sigma/\sigma_0, \phi/\phi_c$).
  The lower panel shows the evolution after the end of inflation of
  $H/H_0$, $\phi/\phi_c$ and $\sigma/\sigma_0$, as a function of the
  number of oscillations of the field $\sigma$, $N=\bar m_\sigma
  t/2\pi$.  It is clear that for the first dozens of oscillations, the
  behavior of both fields is extremely chaotic, and the amplitude of
  the fields decreases rather slowly compared to previous cases. These
  figures correspond to the parameters $\lambda=g^2=10^{-6},
  M=10^{-6}\,M_{\rm P}, m=2\times10^{-10}\, M_{\rm P}$.  }
\end{figure}

\begin{figure}[t]
\centering
\hspace*{-4mm}
\leavevmode\epsfysize=6.1cm \epsfbox{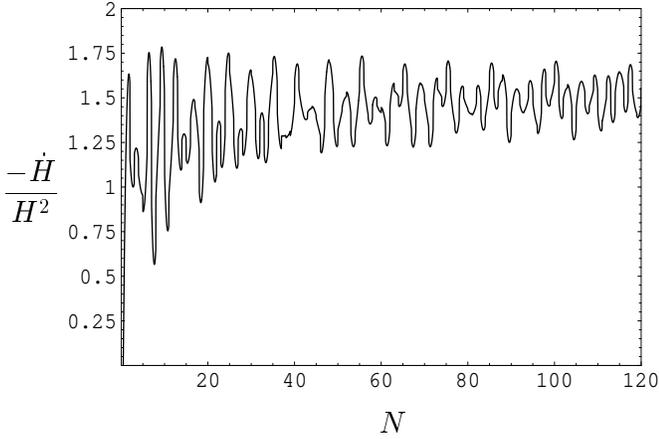}\\[2mm]
\caption[fig7]{\label{fig7} The evolution of the parameter
  $\epsilon=-\dot H/H^2$ after the phase transition in the case
  $g^2=\lambda$. The dust dominated era, $\epsilon=3/2$, is approached
  rather chaotically after many field oscillations, although in less
  than one $e$-fold.}
\end{figure}

In this case, the motion is always two-dimensional in the $(\sigma,
\phi)$ field space: the field $\phi$ oscillates around zero while the
field $\sigma$ oscillates around $\sigma_0$, with similar amplitudes
but varying frequencies, see Fig.~\ref{fig6}. Since the fields are
coupled, their effective masses also vary. This induces the chaotic
behavior of the fields shown in Fig.~\ref{fig6}.\footnote{For a general
discussion of chaos  in hybrid inflation and in preheating, respectively, see
Refs. \cite{Maeda} and \cite{KLS2,bass}.} We have taken as parameters,
$\lambda = g^2 = 10^{-6}, M = 10^{-6}\,M_{\rm P}, m = 2\times10^{-10}
\,M_{\rm P}$. The amplitude of the fields takes many oscillations to
decrease significantly, e.g. after $N=50$, the amplitude of the $\phi$
field is still of order $1/2$, although after one $e$-fold
(corresponding here to $N\simeq340$ oscillations of the $\sigma$
field) it decays like $1/t$ with amplitude $\Phi_0\sim 1/5$.
Meanwhile its frequency is rather chaotic for the first $e$-fold, as one
can appreciate in Fig.~\ref{fig6}, but eventually regularizes and
becomes oscillatory, while the rate of expansion approaches the usual
dust dominated behavior, $\epsilon=-\dot H/H^2 = 3/2$. This can be
appreciated in Fig.~\ref{fig7}.

Even though initially the chaotic behavior of the fields' oscillations
prevents any particle production, as soon as their behavior becomes
regular (after about 1 $e$-fold) the amplitude of the oscillations is
large enough (thanks to the fact that it has not decreased significantly
during the chaotic motion) that one may have some parametric resonance.

Like in the previous case $(g^2\gg\lambda)$, a new field $\chi$ could
couple to $\sigma$ and $\phi$ with potential (\ref{couplings}). Since
$\sigma$ acquires the vacuum expectation value $\sigma_0$, we have to
suppress the coupling to $\sigma$ in order to induce parametric
resonance of $\chi$-particles. In this case, since $g^2=\lambda$, the
condition (\ref{hcond}) becomes $h_1\gg h_2$. Eventually the amplitude
of the field $\phi$ oscillations will decrease to $\Phi(t) < h_2/h_1$
and the explosive production will end. The Mathieu equation is the
same as in the previous case, see Eq.~(\ref{Mchi}), with occupation
number~(\ref{occnum}) and frequency~(\ref{omega}).

\begin{figure}[t]
\centering
\hspace*{-3.5mm}
\leavevmode\epsfysize=6.cm \epsfbox{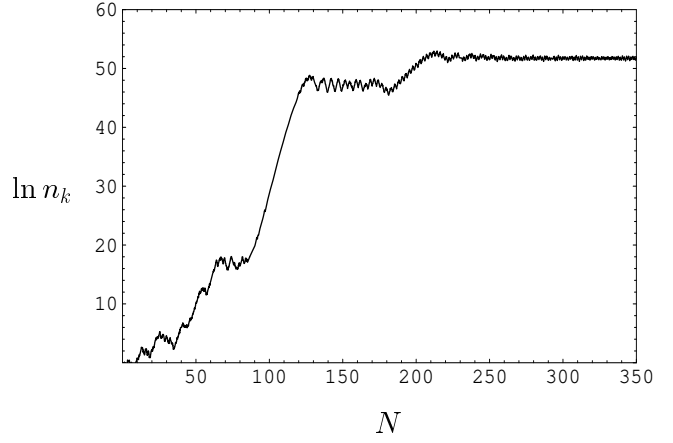}\\[2mm]
\caption[fig8]{\label{fig8} The exponential growth of the occupation
  number $n_k$ of $\chi$-particles as a function of the number of
  oscillations of the field $\sigma$, $N=\bar m_\sigma t/2\pi$, for
  $k=M$ and the parameters of Fig.~\ref{fig6}. One can distinguish
  here the chaotic plateau followed by the broad stochastic resonance
  regime. It acquires a typical growth parameter $\mu \simeq 0.03$
  during the last stages of preheating.}
\end{figure}

We have taken in this case $h_2\simeq0$ and $h_1=4\times10^{-3}$, and
found explosive production of $\chi$-particles, see Fig.~\ref{fig8}.
Note the initial plateau, as the chaotic behavior of $\phi$ prevents a
significant production of $\chi$-particles. Eventually, the stochastic
resonance regime begins, but the growth index $\mu_k\simeq 0.03$
associated with the mode $k=M$ at the last stages of reheating is not
as large as in the previous cases.

Backreaction of the produced $\chi$-particles on the oscillations
of the $\phi$-field will become important at time $t_1$, given by
Eq.~(\ref{N1}), when the expectation value of the $\chi$-field
becomes
\begin{eqnarray}\label{breact}
\langle\chi^2\rangle_1 \simeq {\bar m_\phi^2\over h_1^2} =
{M^2\over h_1^2}\,,
\end{eqnarray}
and the field $\phi$ acquires a large mass
$\sim h_1^2\langle\chi^2\rangle_1$. Particle production still
continues until the effective $q$ parameter enters the narrow
resonance regime, $q\lesssim 1/4$~\cite{KLS2}.

Symmetry restoration, if it exists, depends only on the couplings
of $\chi$ to $\phi$ and $\sigma$, see Eq.~(\ref{srcond}),
\begin{equation}\label{src}
q_1^{1/2} > {h_1^2\over4h_2^2} \gg 1\,.
\end{equation}
Although difficult, this condition could still be satisfied.

In this special case, there is also a small production of
$\sigma$-particles, due to the oscillations of the homogeneous
$\sigma$ and $\phi$-fields. The effective mass (\ref{msigma}) of the
$\sigma$-fluctuations in this case is
\begin{eqnarray}\label{tildemsigma2}
m^2_\sigma &=& \bar m^2_\sigma + g^2\phi^2 +
3\lambda \,(\sigma^2-\sigma_0^2) \,,\\ \nonumber
&=& \bar m^2_\sigma + M^2\Phi^2\sin^2\bar m_\phi t \\
&& \hspace{6mm} + \,3M^2\Big(2\Sigma\sin\bar m_\sigma t +
\Sigma^2\sin^2\bar m_\sigma t\Big) \,,
\end{eqnarray}
where we have used Eqs.~(\ref{Phi}) and (\ref{Sigma})

\begin{figure}[t]
\centering
\hspace*{-3.5mm}
\leavevmode\epsfysize=6.cm \epsfbox{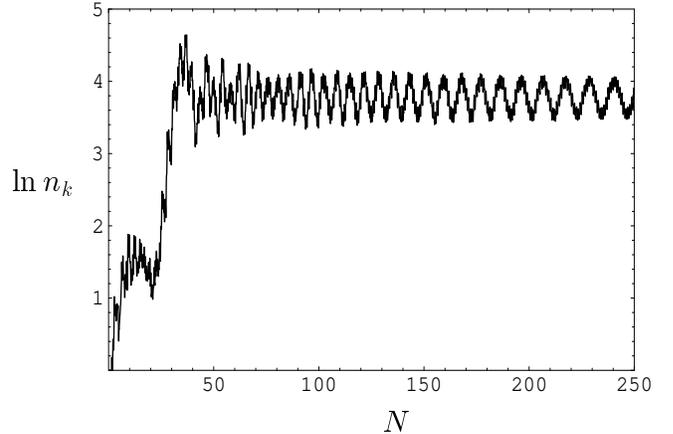}\\[2mm]
\caption[fig8]{\label{fig9} The exponential growth of the occupation
  number $n_k$ of $\sigma$-particles as a function of the number of
  oscillations of the field $\sigma$, $N=\bar m_\sigma t/2\pi$, for
  $k=M$ and the parameters of Fig.~\ref{fig6}. One can only
  distinguish here the narrow resonance regime.  It acquires a typical
  growth parameter $\mu \simeq 0.01$ during the last stages of
  preheating.}
\end{figure}

Since $\bar m^2_\phi = g^2M^2/\lambda \simeq M^2$ and $\bar m^2_\sigma
= 2M^2$, the effective mass (\ref{tildemsigma2}) oscillates with
various frequencies. In this case, we cannot neglect the terms
proportional to $\Phi^2$ and $\Sigma^2$, since they are not small and
represent the oscillations of the two background fields. Initially,
the motion is chaotic and no parametric resonance is possible.
Eventually, oscillations become more regular, when the amplitude of
oscillations is still large, $\Sigma\sim1/2$ and $\Phi\sim1/3$. Then
the Mathieu equation for the production of $\sigma$ particles becomes
\begin{equation}\label{Msigma3}
\psi_k'' + \left[A_\sigma(k) + 2q_\sigma \sin 2z\right]\psi_k = 0\,,
\end{equation}
where $z=\bar m_\sigma t/2$ and
\begin{eqnarray}\label{Asigma3}
A_\sigma(k) &=& {4k^2\over a^2\bar m_\sigma^2} + 4 \,,\\
\label{qsigma3}
q_\sigma &=& {12M^2\over\bar m_\sigma^2}\,\Sigma\,.
\end{eqnarray}
Initially we can neglect backreaction, and $\bar m_\sigma^2 =
2M^2$. In this case $q_\sigma = 6\Sigma_0 \simeq 3$, see
Fig.~\ref{fig6}b. There will be a slight production of
$\sigma$-particles, as can be seen in Fig.~\ref{fig9}a. However, the
effective growth parameter $\mu_k \simeq 0.01$ is never large enough
to dominate the decay of the $\sigma$-field. Backreaction will be
dominated by the $\chi$-particle production, certainly not by
$\sigma$-particle production.

The existence of the regime of chaotic oscillations may be important
in application to realistic models of hybrid inflation. For example,
recently it was argued that in hybrid inflation scenario based on the
$SU(5)$ model, spontaneous symmetry breaking has a tendency to occur
in the direction with a ``wrong'' type of symmetry breaking
$SU(4)\times U(1)$, which was considered as a problem of this
scenario~\cite{Dvali}. However, analogous result is true even for the
standard high temperature phase transitions in the
$SU(5)$~\cite{Kuzmin}, where the phase transition is typically first
order, and the bubbles of the phase $SU(4)\times U(1)$ form much more
often than the bubbles of the phase $SU(3)\times SU(2)\times
U(1)$~\cite{L1981}. Still, the universe may eventually end up in the
proper vacuum state $SU(3)\times SU(2)\times U(1)$, after the bubbles
of the phase $SU(3)\times SU(2)\times U(1)$ ``eat'' the bubbles of all
other phases~\cite{L1981,book}. Similarly, the phase transition in the
$SU(5)$ version of the hybrid inflation model may be first order, with
several different phases being produced. Then one should study the
equilibrium between these phases. But even if the phase transition is
second order, as in the simplest model considered above, one may
wonder whether the chaotic behavior of the scalar fields $\phi$ and
$\sigma$ may split the universe into many domains with different types
of symmetry breaking. Then one should study which of these domains
will eventually survive. This process becomes even more complicated if
one studies backreaction of $\chi$ particles produced during
preheating on the shape of the effective potential. The best way to
investigate this issue is to use lattice
simulations~\cite{Tkachev,TKKL}.

\section{Reheating in hybrid models with two stages of inflation}

So far we have studied the generic behavior of preheating in various
hybrid inflation models. We see that in some of them it is indeed
possible to produce large amounts of particles in an explosive way and
thus reheat the universe rather efficiently.  In all models which we
studied so far inflation ends at the moment when the field $\phi$
reaches its critical value $\phi_c$ and a phase transition with the
generation of the classical field $\sigma$ occurs.

However, this is not the most general situation. There are some hybrid
inflation models where inflation ends before the phase transition, see
e.g. Ref.~\cite{shafi,LR}. There is also an interesting class of models
where inflation continues for a while after the phase
transition~\cite{Guth,GBLW}. This happens if the symmetry breaking
field has a vacuum expectation value $\sigma_0 \sim M_{\rm P}$, and a
mass of order $M\sim1$~TeV.  This requires an extremely small
parameter $\lambda\ll g^2$. As a consequence, after the phase
transition at $\phi=\phi_c$, the symmetry breaking field slowly rolls
down its potential while still inflating the universe, until it starts
oscillating around $\sigma_0$, ending inflation. The amplitude of
oscillations is very small compared with $\sigma_0$, i.e. $\Sigma \ll
1$.

As we already discussed in Section \ref{small}, in this case
preheating is inefficient unless there exists a field $\chi$ strongly
coupled to $\sigma$, which, however, does not acquire mass greater
than $M$ after spontaneous symmetry breaking. It is especially
difficult to arrange it in the case under consideration, because the
amplitude of spontaneous symmetry breaking is $\sim M_{\rm P}$, which
is 16 orders of magnitude greater than $M$. Thus one may expect that
even a very weak interaction of the field $\sigma$ with $\chi$ will
make the field $\chi$ too heavy. In Section \ref{small} we mentioned
that it is possible to have light particles which are strongly coupled
to the field $\sigma$, but it requires either fine tuning or
introduction of flat directions. We will not discuss this possibility
here.

Now let us consider a possibility of a perturbative decay of the
oscillations of the field $\sigma$. In general, this process could be
efficient if this field has a renormalizable interaction with some
light fields, with a large coupling constant $\alpha^2$. However,
there are two problems associated with this possibility. First of all,
as we just mentioned, such an interaction typically makes the fields
interacting with the field $\sigma$ extremely heavy, with a mass $\sim
\alpha \sigma_0 \sim \alpha M_{\rm P}$. Such particles cannot be
produced by perturbative decay of the particles with mass $M\sim1$~TeV
unless $\alpha$ is extremely small, $\alpha < 10^{-16}$. But in this
case the decay rate is extremely small too.

However, even if one finds a way to have some light particles strongly
coupled to the field $\sigma$, its decay still occurs very slowly.
Indeed, suppose that the decay rate $\Gamma$ is very large, $\Gamma
\sim M$.  Then naively one could expect that the field decays within
the time $\Delta t \sim \Gamma^{-1} \sim M^{-1}$, which is of the same
order of magnitude as the Hubble constant at the end of inflation in
this model. In such a case the energy density of the oscillating
scalar field $\rho_\sigma \sim M^2\sigma_0^2 \sim M^2 M_{\rm P}^2$
would be rapidly converted to the thermal energy $\sim T^4$, which
would heat the universe up to the temperature $T_r \sim 10^{-1}\sqrt
{M M_{\rm P}} \sim 10^7 M$~\cite{KLS2}.  However, this is hardly
possible. Indeed, strongly interacting particles in thermal
equilibrium typically acquire effective mass $\sim T$. As soon as this
mass becomes comparable to $M$, reheating shuts down.

As a result, instead of rapidly reaching the temperature $T_r \sim
10^{-1}\sqrt {M M_{\rm P}}$, the universe remains at the nearly
constant temperature $T \sim M \sim 1$ TeV. The decay of the field
$\sigma$ continues for a very long time, until the energy density of
this field drops down from $M^2 M_{\rm P}^2$ to $M^4$; see Ref.
\cite{ADL85} for a discussion of a similar effect.  Thus, reheating
temperature in this model may be very small; reheating may end just
before the electroweak phase transition or even after it.  This may
have important implications for the theory of baryogenesis and for the
cosmological moduli problem.

But the situation may be even more interesting and complicated.
Indeed, as was noted in Ref.~\cite{Guth}, the existence of a second
stage of inflation in this model may produce a very narrow peak in the
spectrum of density perturbations. A further investigation of this
issue have shown that this peak is in fact much higher than originally
expected, see Ref.~\cite{GBLW}.  The density perturbations
corresponding to this peak may lead to a copious black hole production
even if the universe after inflation rapidly enters the radiation
dominated stage. However, this effect becomes even much more
pronounced if the universe for a long time remains matter dominated,
because without radiation pressure there is nothing to prevent
gravitational collapse. Therefore even if the peak in spectrum of
density perturbations is not very high, it may still lead to a copious
black hole formation.

If our analysis of reheating in this model is correct, then reheating
completes at a time which is at least 16 orders of magnitude longer
than the age of the universe at the end of inflation. All this time
the universe remains matter dominated.  In this case we expect a huge
number of black holes with a narrow range of masses to be produced
soon after inflation, so that most of the matter after inflation will
be in the form of primordial black holes. If the second stage of
inflation is prolonged, the masses of the black holes will be
extremely large, which will lead to unacceptable cosmological
consequences.  However, for a wide range of the parameters the second
stage of inflation will be relatively short.  In such a case the black
hole masses will be very small, and they will evaporate before
primordial nucleosynthesis. The peak in the spectrum of density
perturbations in this model is very narrow. As a result, all black
holes will have mass of the same order of magnitude. Evaporation of
these black holes will happen almost simultaneously, at the time
related to their mass.  This process will produce all the radiation
and matter we see in the universe today~\cite{GBLW}.

\section{Conclusions}

In this paper we considered the initial stages of reheating in the
hybrid inflation scenario. We have found that in certain cases this
process begins with a stage of parametric resonance.  In such a
situation the standard perturbative approach to the theory of
reheating should be considerably modified. But even in the situations
where preheating is inefficient, a detailed investigation of the
behavior of the coupled system of two fields $\phi$ and $\sigma$ is
necessary for a proper investigation of the perturbative reheating
regime. For example, as we have shown, for $\lambda \gg g^2$ all
energy after inflation is stored in the oscillations of the field
$\phi$, so one should study perturbative or nonperturbative decay of
this field. For $\lambda \ll g^2$ all energy is stored in the
oscillations of the field $\sigma$, so one should study its decay.
Meanwhile for $\lambda \sim g^2$ the fields enter a regime of chaotic
oscillations where energy density is transferred back and force
between the fields $\phi$ and $\sigma$.  Investigation of this regime
may be interesting not only for the investigation of reheating, but
also for the study of symmetry breaking pattern in realistic models of
hybrid inflation. We have found that preheating can be very efficient
if the effective masses of the fields $\phi$ and $\sigma$ are much
greater than the Hubble constant at the end of inflation, or if these
fields are coupled to other light scalar (or vector) fields~$\chi$.

In addition to the simplest hybrid inflation scenario~\cite{hybrid},
we also studied a recently proposed scenario with two stages of
inflation~\cite{Guth,GBLW}. Rather unexpectedly, we found that
preheating, as well as the standard mechanism of reheating, in this
scenario is extremely inefficient. In such a situation, reheating may
occur in a very unusual way, via formation of
primordial black holes and their subsequent evaporation.  Even though
this process may seem somewhat exotic, it may be the leading mechanism
of reheating in inflationary models with a peak in the spectrum of
density perturbations~\cite{GBLW}, as well as in a much more general
class of hybrid inflationary models with a blue spectrum of density
perturbations.

\section*{Acknowledgements}

J.G.B. thanks G. Dvali and J.R. Espinosa for useful discussions. He
is also grateful to the Physics Department at Stanford University, for
their hospitality during part of this work. A.L. is very grateful to
L. Kofman and I. Tkachev for the discussions of the theory of
preheating. The work of A.L. was supported by NSF grants PHY-9219345 and
AST95-29225.
This work was also supported by a NATO Collaborative Research Grant,
Ref.~CRG.950760.

\end{document}